\documentclass[aps, prb,twocolumn, showpacs]{revtex4}
\usepackage{graphicx}
\usepackage{dcolumn}
\usepackage{amsmath}

\begin{document}

\title{Longitudinal and spin-Hall conductance \\ of a two-dimensional Rashba system
with arbitrary disorder}
\author{C\u at\u alin Pa\c scu  Moca}
\affiliation{
Department of Physics, University of Oradea, 410087 Oradea, Romania \\
Department of Physics and Astronomy, Clemson University, 29634, Clemson.
}
\author{D.\ C.\ Marinescu }
\affiliation{
 Department of Physics and Astronomy, Clemson University, 29634, Clemson.
}

\date{\today}
\begin{abstract}
We calculate the longitudinal and spin-Hall conductances in
four-lead bridges with Rashba - Dresselhaus spin-orbit interactions.
Numerical results are obtained both within Landauer-B\" uttiker
formalism and by the direct evaluation of the Kubo formula. The
microscopic Hamiltonian is obtained in the tight-binding
approximation in terms of the neareast-neighbor hopping integral
$t$, the Rashba spin-orbit coupling $V_R$, the Dresselhaus
spin-orbit coupling $V_D$ and an Anderson-like, on-site disorder
energy strength $W$. We reconfirm that below a critical disorder
threshold, the spin-Hall effect is present. Further, we study the
effect on the two conductivities of the Fermi energy,
Rashba/Dresselhaus coefficient ratio, and system size.
\end{abstract}

\pacs{72.23.-b, 72.10.-d, 72.15.Gt}
\maketitle

\section{Introduction}

Known to exist for a long time \cite{general}, the spin-orbit (SO)
coupling in two dimensional electronic systems (2DEG) has received a
lot of attention lately motivated by its potential applications in
spintronics. Recent experiments\cite{SdH} have demonstrated that the
magnitude of the spin-orbit coupling can be modified by a voltage
gate, hence generating the premise of the possible manipulation of
spin currents by electric fields alone. The two sources of the
spin-orbit coupling are the inversion asymmetry of the confining
potential in the direction perpendicular to the 2DEG (Rashba) and
the bulk asymmetry and interface inversion asymmetry
(Dresselhaus)\cite{Dresselhaus}.

In a very interesting development\cite{Sinova}, Sinova {\it et. al}
predicted that a spin-Hall current of transverse spin component
appears in a 2DEG with SO coupling as a response to a in-plane
electric field. This spin current has a universal value, equal to
$e/8\pi$. The intrinsic spin-Hall effect is quite different from
the extrinsic spin-Hall effect\cite{Hirsch} proposed by Hirsch,
which is generated by impurity scattering. The possible existence
and persistence in disordered systems of the intrinsic spin-Hall
effect (SHE) have been the focus of many recent papers\cite{recent}.
The question of whether arbitrary small amounts of disorder suppress
or not the intrinsic SHE is still awaiting a definite answer. Some
analytical calculations \cite{Zhang} claim that SHE does not survive
even in the weak disorder regime, while others \cite{Murakami} provide
arguments that SHE is robust and weak disorder in the system is not
enough to destroy this effect.  While the problem was studied in
more detailed using analytical methods, there are few unbiased
numerical calculations\cite{Nikolic} at present.

In this work, we present numerical results for the longitudinal and
spin-Hall conductivities of a 2DEG with spin-orbit interactions,
both Rashba and Dresselhaus, in the presence of disorder. These
values are obtained within a spin-dependent Landauer-Buttiker
formalism, developed for a microscopic Hamiltonian written in a
tight-binding approximation that incorporates both the spin-orbit
interaction and disorder. As a further check, we calculate the same
conductances by using the Kubo formalism and find good agreement
between the two sets of results. Our findings suggest that the
spin-Hall effect occurs in disordered systems, for as long as the
disorder remains below a critical threshold value. We also study the
dependence of the conductivities on the Fermi energy, system size,
and on the relative strengths of the two types of SO coupling.

In the section II of the paper we present the general framework of
spin-dependent LB formalism used for computing the spin-Hall
conductance, while in section III we show and discuss our
results. For comparison, in the appendix, we compute the same conductances
by using the Kubo formalism.

\section{System Description}
The single particle Hamiltonian for an electron of momentum
$\mathbf{p}=(p_x, p_y)$, spin $\boldsymbol{\sigma} =
(\sigma_x,\sigma_y,\sigma_z)$, and effective mass $m^*$, in a 2DEG
with Rashba ($\alpha$) and Dresselhaus ($\beta$) spin-orbit interactions is:
\begin{equation}
H = \frac{{\mathbf p}^2}{2 m^{\star}} +\alpha \left( \sigma _x p_y -\sigma_y p_x\right )
    +\beta \left( \sigma _x p_x -\sigma_y p_y \right )\;. \label{eq:hamiltonian}
\end{equation}
The  relative strengths of the Rashba and Dresselhaus terms,
$\alpha/\beta$ describing the spin-orbit coupling in semiconductor
quantum wells, are available from photocurrent
measurements\cite{Ganichev}. The interplay of the two SO couplings
has been also lately subject to intense theoretical investigations
 with respect to other physical phenomena
such as magneto-oscillation phenomena in quantum wells or 
spin splitting of the electron energy states in quantum dots
\cite{rashba_dressel}.

We discretize the Hamiltonian using a tight-binding approach, where
the solution domain is filled with a regular virtual lattice. The
Hamiltonian is constructed over this lattice assuming only
neareast-neighbor coupling. This can be done straightforwardly by
using the projections on   $x$ and $y$ direction of the momentum
operator ${\mathbf p} =-i\hbar\nabla$ in Eq. (\ref{eq:hamiltonian}).
The resulting tight-binding Hamiltonian is:
\begin{eqnarray}
H &=&  \sum\limits_{i,\alpha,}\varepsilon_i c_{i\alpha}^{\dagger} c_{i\alpha}
-t\sum\limits_{<i,j>,\alpha} c_{i\alpha}^{\dagger}c_{j\alpha}\label{eq:tight_binding_hamiltonian_large}\\
&+& V_R\sum\limits_{i} \left[\left( c_{i\uparrow}^{\dagger}c_{i+\delta_x \downarrow}-
c_{i\downarrow}^{\dagger}c_{i+\delta_x \uparrow}\right)\right . \nonumber \\
& &-i\left. \left( c_{i\uparrow}^{\dagger}c_{i+\delta_y \downarrow}+
c_{i\downarrow}^{\dagger}c_{i+\delta_y \uparrow}\right) \right]\nonumber\\
&+& V_D\sum\limits_{i} \left[(-i)\left( c_{i\uparrow}^{\dagger}c_{i+\delta_x \downarrow}+
c_{i\downarrow}^{\dagger}c_{i+\delta_x \uparrow}\right)\right . \nonumber \\
& &+\left. \left( c_{i\uparrow}^{\dagger}c_{i+\delta_y \downarrow}-
c_{i\downarrow}^{\dagger}c_{i+\delta_y \uparrow}\right)
\right]\;.\nonumber
\end{eqnarray}
Here $t=\hbar^2/(2m^{\star} a_{0}^{2})$ is the hopping integral,
$V_R=\hbar \alpha / a_0$ and $V_D=\hbar \beta /a_0$ are the Rashba
and Dresselhaus coupling strengths, respectively, renormalized by
the lattice constant $a_{0}$, and $\delta_x$ and $\delta_y$ are the unit
vectors along the $x$ and $y$ directions. The hopping matrix element
$t$ represents the unit of energy in our calculations. The second,
third, and last terms in Eq.~(\ref
{eq:tight_binding_hamiltonian_large}), can be combined and a compact
expression for the Hamiltonian can be written in the form:
\begin{eqnarray}
H &=&  \sum\limits_{i,\mu}\varepsilon_i c_{i\mu}^{\dagger} c_{i\mu}
-\sum\limits_{<i,j>,\mu, \nu}t_{ij}^{\mu\nu} c_{i\mu}^{\dagger}
c_{j\nu}\;,\label{eq:tight_binding_hamiltonian}
\end{eqnarray}
where $c_{j\mu}$ $(c_{j\mu}^{\dagger})$ is the annihilation
(creation) operator of an electron of spin index $\mu$ at site $j$.
The first term in Eq. (\ref{eq:tight_binding_hamiltonian}) is the
on-site disorder, as in the Anderson model, with $\varepsilon_i$, a
random energy generated by a box distribution $\varepsilon_i\in
[-W/2, W/2]$. The SO interactions are directly incorporated in the
hopping term which acquires position and spin dependence.
\begin{figure}[h]
\begin{center}
\includegraphics[width=2.0in]{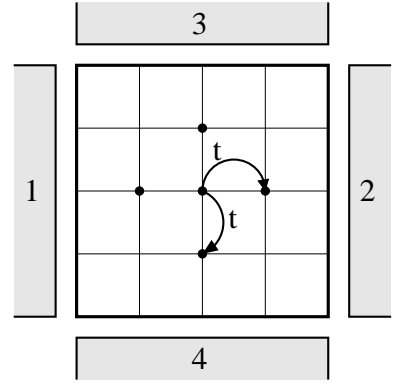}
\end{center}
\vspace*{0.05cm} \caption{Graphical depiction of the lattice model used for
computing the spin-Hall conductance. Four metallic leads
(represented as the dashed regions) acting as  injector (1),
detector (2) and voltage probes (3 and 4) are
 are attached to the 2DEG. Position and spin
dependence are not explicitly decided  for the hopping integral. }
\label{fig:lattice}
\end{figure}
The Hamiltonian given by Eq.~(\ref{eq:tight_binding_hamiltonian}) is
studied in a N$\times$N square lattice, as presented in Figure
\ref{fig:lattice}. Each metallic lead attached to the sample is
considered a perfect semi-infinite wire, without disorder and SO
interactions. $V_R$ and $V_D$ are also assumed to be zero in leads 3
and 4 in order to avoid spin flips at the boundaries. Throughout our
calculations we use the same values for the cross sections of leads
and sample, in order to eliminate scattering induced by the
wide-to-narrow geometry \cite{Szafer}.

Within the LB formalism the total current in terminal $p$ is given
by $I_{p} =e^2/h\sum_{q\ne p}T_{pq}(V_p-V_q) $ where the sum is over
all the other leads $q$ connected to the system. Spin current can be
defined in a similar way, up to a constant: $I_{p,\mu}^{spin} =
e/(4\pi)\sum_{q\ne p, \nu } T_{pq}^{\mu\nu}(V_p-V_q)$.  The voltages
are computed by considering ballistic transport between all the
connected terminals and imposing the following  boundary conditions:
$V_2=0$ (fixes the arbitrary zero of voltage), $I_3 =
2e/\hbar\sum_{\alpha}I_{3, \alpha}^{spin} =0$, $I_4 =
2e/\hbar\sum_{\alpha}I_{4,\alpha}^{spin} =0$ (terminals  $3$ and $4$
are voltage probes) and  $I_1+I_2 =0$, (guarantees that current
flows  between terminals $1$ and $2$). The zero temperature
conductance, $\mathbf G$, that describes the spin-resolved transport
measurements, is related with the transmission matrix $\mathbf T$,
as in :
\begin{equation}
{\mathbf G}= \frac{e^2}{h}{\mathbf T}=\frac{e^2}{h}\left (
\begin{array}{cc}
T^{\uparrow \uparrow} & T^{\uparrow \downarrow} \\
T^{\downarrow \uparrow} & T^{\downarrow \downarrow}
\end{array}
\right ),\label{eq:t_matrix}
\end{equation}
[Indices $p$ and $q$ were suppressed in writing Eq.
(\ref{eq:t_matrix})]. $T_{pq}^{\mu\nu}$ represents the transmission
probability over all the conduction channels to detect a spin $\mu$
in the lead $p$ arising from an injected spin $\nu$ electron in lead
$q$, when both spin-flip and non-spin-flip processes are considered.
The transmission coefficient can be calculated as $T_{pq}^{\mu\nu} =
Tr[\Gamma_p^{\mu}G_R \Gamma_q^{\nu}G_A]$ where $\Gamma_p^{\mu} =
i(\Sigma_{p}^{\mu}-\Sigma_p^{\mu\dagger})$ with $\Sigma_{p}^{\mu}$
the retarded self-energy due to the interaction between the sample
and the lead for spin-channel $\mu$. The self-energy contribution is
computed by modeling each terminal as a semi-infinite perfect wire.
In our  tight-binding model, the hopping between the lead orbitals
and between the leads and the sample orbitals are equal\cite{Datta}
with $t$ (unit of energy). The self-energy matrix, which is diagonal in spin indices,
can be written as:
\begin{equation}
\Sigma_p = \left(
\begin{array}{cc}
\Sigma_p^{\uparrow} & 0\\
0 & \Sigma_p^{\downarrow}
\end{array}
\right)
\end{equation}
with $\Sigma_p^{\uparrow}=\Sigma_p^{\downarrow}$ for a perfect
metallic lead. The retarded Green's function is computed as $G_R =
(E_{F}-H-\sum_{p=1}^4\Sigma_p )^{-1}$, where $E_{F}$ is the Fermi
energy and $H$ is the Hamiltonian in
Eq.~(\ref{eq:tight_binding_hamiltonian}). The advanced Green's
function is, of course, $G_A = G_R^{\dagger}$.

In the LB formalism, the total scattering  between two lead $p$ and
$q$ can be simply written as the sum over all spin components
$T_{pq} = T_{pq}^{\uparrow\uparrow} +T_{pq}^{\uparrow\downarrow}
+T_{pq}^{\downarrow\uparrow}+T_{pq}^{\downarrow\downarrow}$. Two
other useful combinations\cite{Souma} are $T_{pq}^{in} =
T_{pq}^{\uparrow\uparrow} +T_{pq}^{\uparrow\downarrow}
-T_{pq}^{\downarrow\uparrow}-T_{pq}^{\downarrow\downarrow}$ and
$T_{pq}^{out} = T_{pq}^{\uparrow\uparrow}
+T_{pq}^{\downarrow\uparrow}
-T_{pq}^{\uparrow\downarrow}-T_{pq}^{\downarrow\downarrow}$.
$T_{pq}^{out}$ represents the difference between the  transmission
probabilities
 to detect an electron in the lead $p$ arising from an injected spin
$\uparrow$ ($\downarrow$) electron in lead $q$. These expressions
allow us to compute the spin-Hall conductance, as
\begin{equation}
G_{sH} = \frac{I_{3,\uparrow}^{spin}-I_{3,\downarrow}^{spin}}{V_1}\label{eq:spin_hall_conductance}
\end{equation}
Finally, by using the voltages  obtained inverting the multiprobe
equations, the spin-Hall conductance becomes:
\begin{equation}
G_{sH} = e/(8\pi)(T_{13}
^{out}+T_{43}^{out}+T_{23}^{out}-T_{34}^{in}-2T_{31}^{in})\;.
\end{equation}

At the same time, the longitudinal conductance, $G_L =I_2/(V_1 -
V_2)$, is written as:
\begin{equation}
G_L=e^2/h\left( T_{21} +0.5\, T_{32} +0.5\, T_{42} \right)\;,
\end{equation} when four terminals are connected to the sample as in Figure
\ref{fig:lattice}.
The spin-Hall and longitudinal conductances are the central
quantities of our analysis. In the next section, we present results
showing their dependence on the Fermi energy, system size, and
disorder strength.

\section{Results and Discussion}
\begin{figure}[tbh]
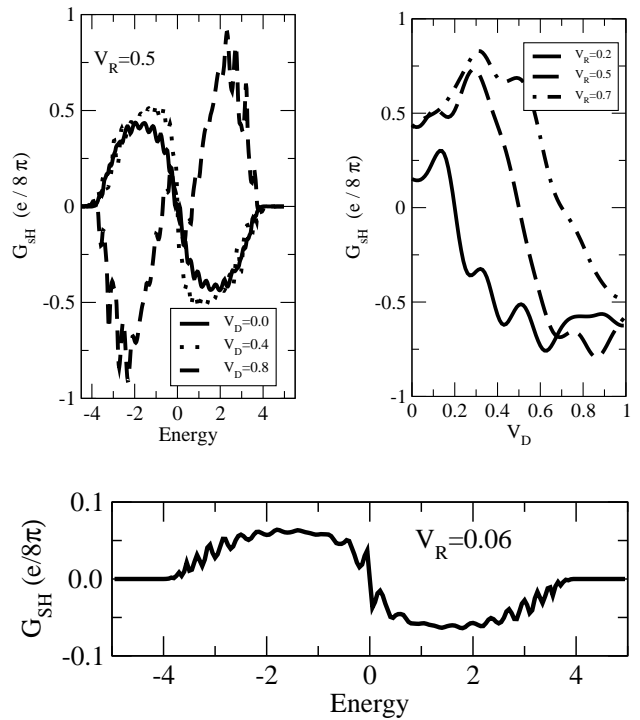

\begin{minipage}[t]{0.47\linewidth}
\centering
\includegraphics[width=1.5in]{spin_hall_conductance_disorder_0.0.eps}
\end{minipage}\hspace {0.1in}
\begin{minipage}[t]{0.47\linewidth}
\centering
\includegraphics[width=1.5in]{spin_hall_dressel_disorder_0.0.eps}
\end{minipage}
\vspace {0.1in}
\begin{center}
\includegraphics[width=3.2in]{spin_hall_conductance_experiment.eps}
\end{center}
\caption{ Upper left panel: The Fermi energy dependence of the
spin-Hall conductance (SHC) of a two dimensional four-probe bridge
in the clean limit, for a fixed Rashba coupling $V_R =0.5$ and for
different Dresselhauss energies. Upper right panel: SHC dependence on $V_D$ for
different Rashba couplings, for Fermi energy $E_F=-2 t$ in the clean
limit.
 For $V_R=V_D$, the SHC vanishes.
The system size is $20\times 20$. Lower panel: The SHC represented
as function of $E_F$ for $V_R =0.06$ and $V_D=0.0$.}
\label{fig:spin_hall_clean_limit}
\end{figure}

{\bf Clean limit}. The clean limit dependence of the spin-Hall
conductance (SHC) on the Fermi energy is shown in Figure
\ref{fig:spin_hall_clean_limit}. The electron-hole symmetry is
preserved throughout the calculation, so the SHC vanishes at the
band center $E_F=0$ and is an odd-function relative to the Fermi
energy, in agreement with the results of Ref [\onlinecite{Nikolic}]. 
The small oscillations observed in the energy dependence are finite size
effects related with the discontinuities in the self-energy
contribution from the terminals and with the discrete energy levels.

\begin{figure}[h]
\begin{center}
\includegraphics[width=3.0in]{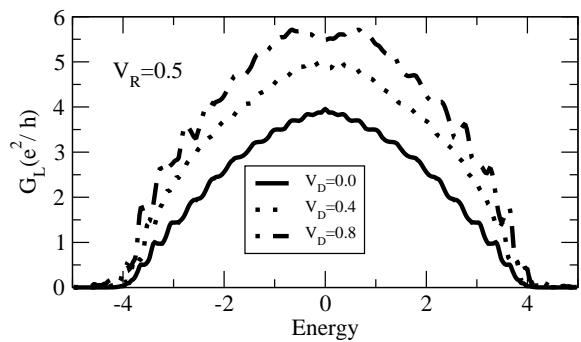}
\end{center}
\vspace*{0.05cm} \caption{The longitudinal conductance as function
of the Fermi energy for different Dresselhaus SO couplings, as
indicated.  The system size is $20\times 20$.}
\label{fig:conductance_disorder_0}
\end{figure}

\begin{figure}[h]
\begin{center}
\includegraphics[width=3.5in]{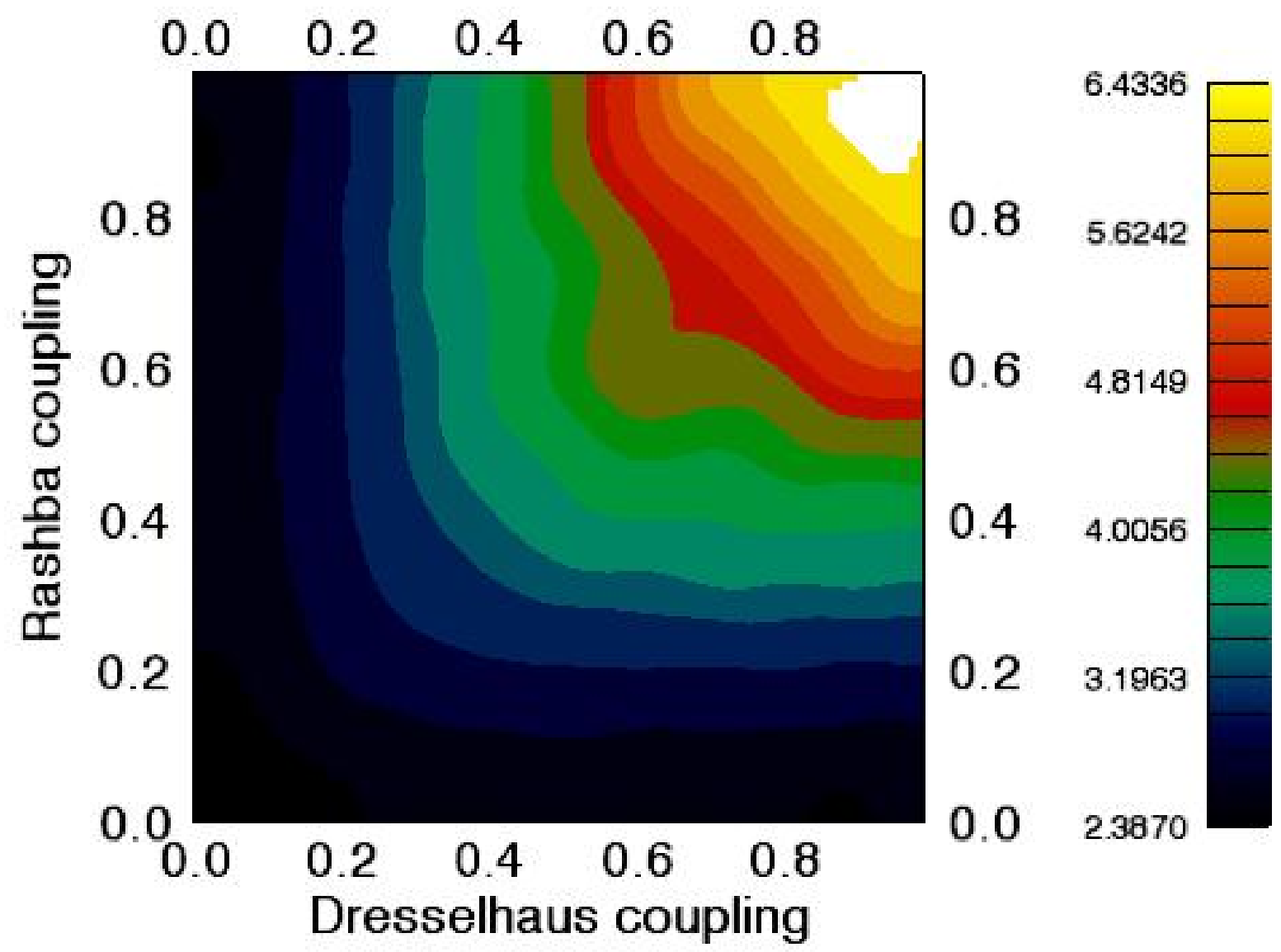}
\end{center}
\vspace*{0.05cm} \caption{(Color online) Longitudinal conductance plotted as a
function of $V_R$ and $V_D$ for a system size $20\times 20$ and for
a Fermi energy $E_F=-2t$. The spectrum is anti-symmetric along the
$V_D=V_R$ line. The spin-Hall conductance is positive for $V_R>
V_D$, negative for $V_R<V_D$ and vanishes for $V_D=V_R$.}
\label{fig:surface}
\end{figure}
\begin{figure}[h]
\begin{center}
\includegraphics[width=3.1in]{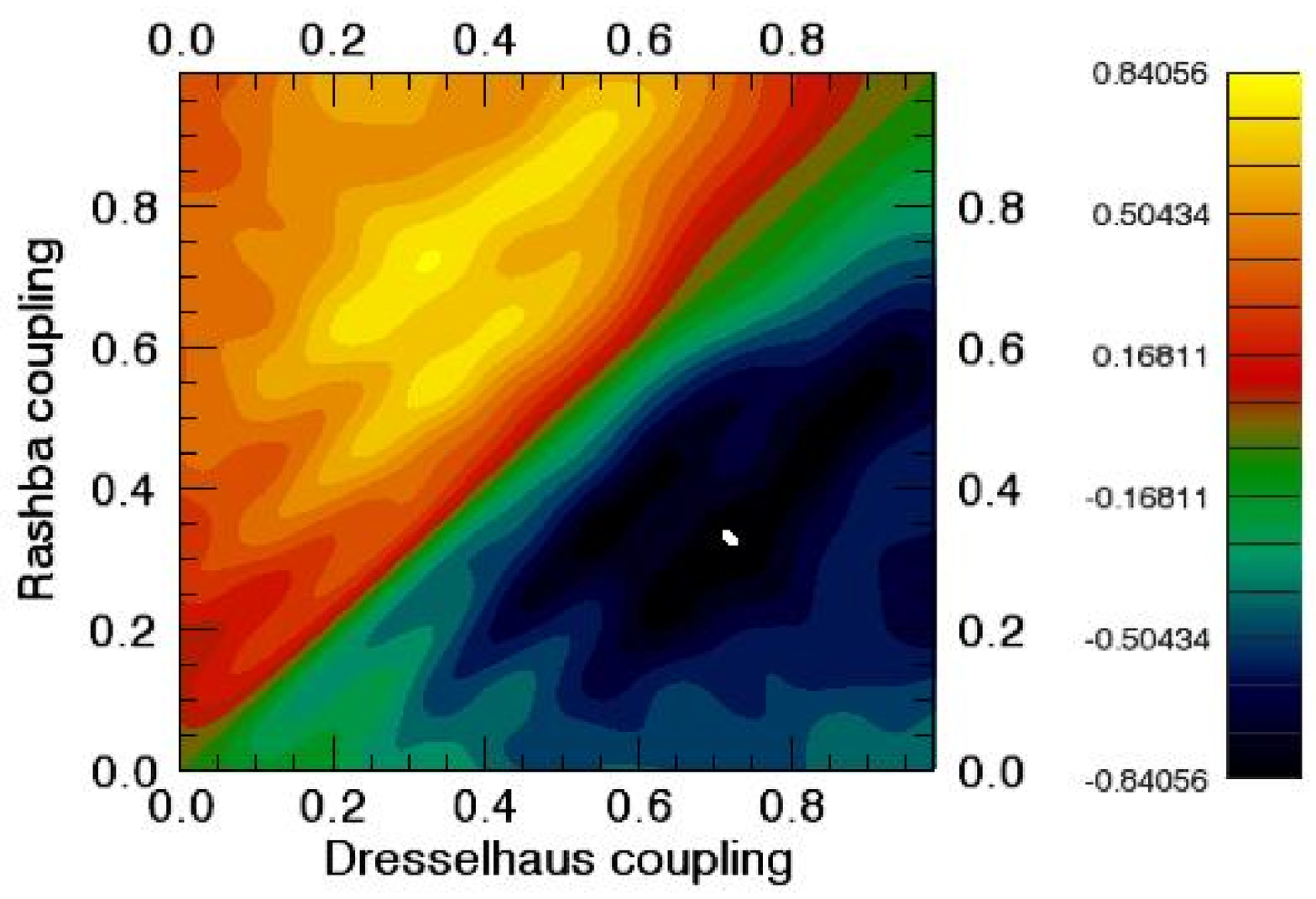}
\end{center}
\vspace*{0.05cm} \caption{ (Color online) The spin-Hall conductance plotted as a
function of $V_R$ and $V_D$ for a system size $20\times 20$ and for
a Fermi energy $E_F=-2t$. The spectrum is anti-symmetric along the
$V_D=V_R$ line. The spin-Hall conductance is positive for $V_R>
V_D$, negative for $V_R<V_D$ and vanishes for $V_D=V_R$.}
\label{fig:spin_hall_surface}
\end{figure}
\begin{figure}[h]
\begin{center}
\includegraphics[width=3.2in]{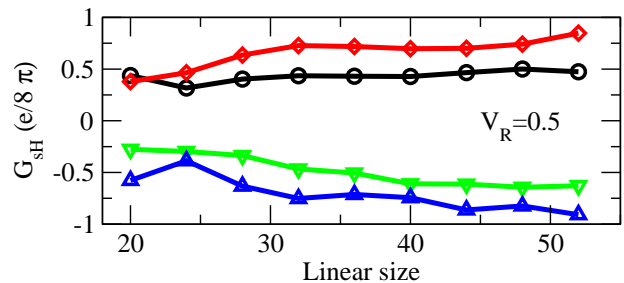}
\end{center}
\vspace*{0.05cm} \caption{ (Color online) Linear system size dependence of the
spin-Hall conductance for a Fermi energy $E_F=2.0 t$, with $V_R=0.5$
and for $V_D$ = \{0.0 $(\circ)$, 0.3 $(\diamond)$, 0.6
$(\bigtriangledown)$, and 0.9 $(\bigtriangleup)$\}. }
\label{fig:scaling}
\end{figure}
Another important parameter  is the ratio $r=V_R/V_D$. When $r=1$,
for any energy, $G_{sH}=0$ (see Fig. \ref{fig:spin_hall_clean_limit}
- right panel,  and Fig. \ref{fig:spin_hall_surface}). For a
hole-like behavior ($E_F<0$) and $r>1$, $G_{sH}$ is positive, while
for $r<1$, $G_{sH}$ changes sign, demonstrating that the spin
current is generated in the direction of the major driving field
\cite{Sinitsyn}. Experimentally\cite{Ganichev, Knap}, the tuning parameter $r$ could be
varied between $1.5$ and $2.5$.

Figure \ref{fig:conductance_disorder_0} presents the effect  of the
Dresselhaus SO coupling on the longitudinal conductance as function
of Fermi energy, for a fixed value of the Rashba coupling. In
contrast, in Fig. \ref{fig:surface} the Fermi energy is fixed to $E_F
= -2t$ and the longitudinal conductance is plotted as function of
both Rashba and Dresselhaus interactions. Here, $r=1$ still
represents a symmetry line in the parameter space ($V_R, V_D$). For
a fixed value of the Fermi energy, we found the symmetry relation:
$G_L(V_R, V_D) =G_L(V_D, V_R)$.

In Figure \ref{fig:spin_hall_surface} we present the spin-Hall
conductance as function of $V_R$ and $V_D$. SHC is anti-symmetric
along the $V_D=V_R$ line. The Fermi energy is fixed at $E_F=-2t$ and
the system size is $20\times 20$. For a lattice parameter of
$a_0=5.0$ nm and electron effective mass $m^{\star}=0.068m$ (in
GaAs), the hopping integral is $t\simeq 19.0 $meV. A typical value
for the Rashba coupling\cite{Miller} is $\sim 50-80$
meV$\cdot$\AA, which corresponds to $V_R=1-1.6$ meV, with a
typical ration $V_R/t\simeq 0.05-0.08$. The results presented in
Fig. \ref{fig:spin_hall_clean_limit} (upper panel) are beyond the
experimental reach. In Figure \ref{fig:spin_hall_clean_limit} (lower
panel) we represent the Fermi energy dependence of the SHC with a
experimental accessible value for the spin-orbit interaction
strength, and, as expected, the SHC amplitude is strongly reduced.
However, the overall behavior is preserved.

The effect of scaling as function of system size is presented in 
Figure~ \ref{fig:scaling}.
Spin Hall conductance is essentially constant  up to, at least, system sizes $50\times 50$.
However we emphasize that the effect of boundaries, due to the
attached leads, may be very important and in principle can hide the true
nature of the bulk spin-Hall effect.

Our analysis shows that in the clean limit, a non-universal value
for SHC exists, in agreement with other numerical
calculations\cite{Nikolic}. SHC strongly depends on the strength of
spin-orbit couplings, while the spin current is always along the
driving field in the system and depends on the relative strength of
the Rashba and Dresselhaus couplings. In the hole-like
(electron-like) regime, characterized by $E_F<0$ ($E_F>0$),
$sgn(SHC)\sim \pm sgn (V_R-V_D)$.

\begin{figure}[tbh]
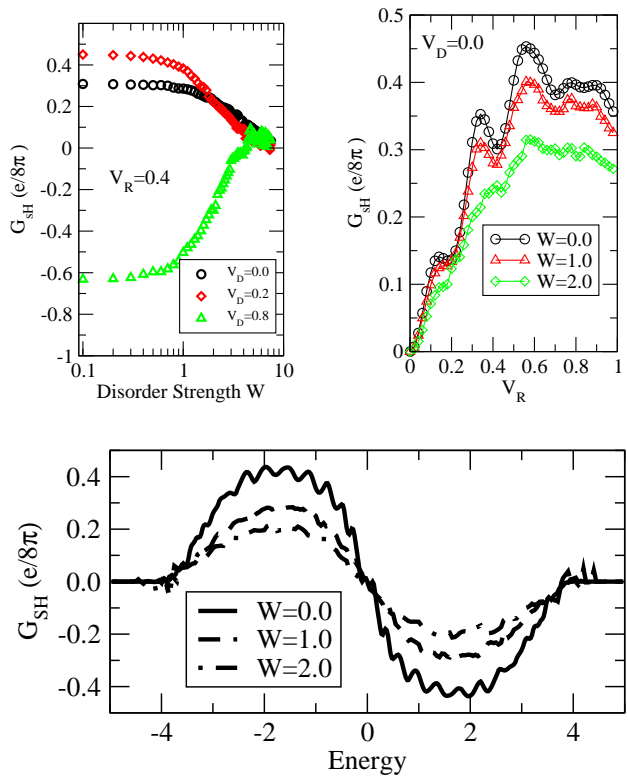

\begin{minipage}[t]{0.47\linewidth}
\centering
\includegraphics[width=1.5in]{disorder_2.eps}
\end{minipage}\hspace {0.1in}
\begin{minipage}[t]{0.47\linewidth}
\centering
\includegraphics[width=1.45in]{rashba_disorder.eps}
\end{minipage}
\vspace {0.1in}
\begin{center}
\includegraphics[width=3.2in]{spin_hall_conductance_disorder.eps}
\end{center}
\caption{ (Color online) Left: Disorder strength dependence of the SHC
with  Rashba spin-orbit strength $V_R=0.4$.
Right: SHC as function of Rashba coupling for different disorder strengths $W$.
Lower panel: SHC as function of Fermi energy and for different disorder strengths $W$.
Electron-hole symmetry is preserved in the presence of disorder.
In the upper panel Fermi energy is $E_F=-2.0 t$. System size is $16 \times 16$ in all figures.}
\label{fig:disorder}
\end{figure}

{\bf Arbitrary disorder}. A system with time reversal symmetry, but
with spin rotational symmetry broken by the spin-orbit coupling,
belongs to the symplectic universality class. It is well established
by now that $SU(2)$ models with chiral symmetry exhibit an Anderson
transition in two-dimension\cite{Ohtsuki}. Critical disorder
strength was estimated to be  $W_C\simeq 5.9$ and the critical
exponent for the localization length $\nu\simeq 2.74$. In our model,
different values for the hopping coupling may lead to different
values for the disorder strength. However, it is understood that SHC
cannot survive in the insulating regime of a 2DEG, because any
localized state cannot contribute to SHC. It is still not clear
whether SHC vanishes in the diffusive transport when the mobility
edge $\pm E_C$ moves towards the band center and localized states in
the band tails coexist with extended states in the band center. To
answer this question we study the effect of disorder on SHC.
 In Figure \ref{fig:disorder} (Left panel) we represent the SHC as function of
disorder strength for different $V_D$. We find that $G_{sH}$ can be
suppressed by a strong scattering when  $W \ge 4-5$, 
close to the metal-insulator transition disorder strength. In the left panel
the Dresselhaus coupling is zero and the Rashba coupling strength
dependence of SHC is presented. For comparison we have plotted also
the result when no disorder is present in the system.

\begin{figure}[tbh]
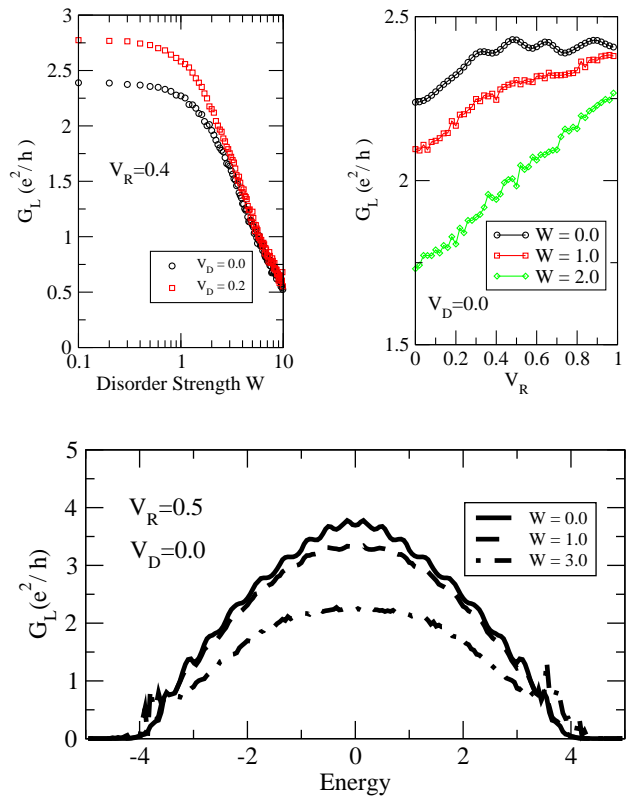

\begin{minipage}[t]{0.47\linewidth}
\centering
\includegraphics[width=1.5in]{longitudinal_disorder.eps}
\end{minipage}\hspace {0.1in}
\begin{minipage}[t]{0.47\linewidth}
\centering
\includegraphics[width=1.45in]{longitudinal_rashba_disorder.eps}
\end{minipage}
\vspace {0.1in}
\begin{center}
\includegraphics[width=3.2in]{conductance_disorder.eps}
\end{center}
\caption{(Color online) Longitudinal conductance as function of disorder strength (upper left panel), and
 as function of Rashba spin-orbit coupling strength (upper right panel)
for different disorder amplitude. Lower panel: Effect of
disorder on the energy dependence of the longitudinal conductance.
The system size is $16\times 16$. Disorder average is over 1000 samples.}
\label{fig:longitudinal_disorder}
\end{figure}
Energy dependence was also considered in the presence of disorder.
(see Figure \ref{fig:disorder}, lower panel).

In the insulating regime, all states are localized, so the absence
of extended states available for transport  at the  Fermi level
leads to a vanishing SHC. When disorder is weak enough, extended
states in the band center coexist with insulating states 
localized mostly in the band tails. These extended states may be
responsible for non-vanishing SHC when small amounts of disorder are
present in the system. 

It is well known that in Landauer-B\" uttiker formalism the attached leads
play an important role, affecting the system self-energy, and can
alter the nature of the bulk spin-Hall effect, while this is not the
case in the Kubo formalism. 

To study the effect of terminals on the spin-Hall and 
longitudinal conductances we  did a direct calculation
of conductances within the Kubo formalism (see  the appendix for further details).
We found good agreement between the 
conductance values obtained in both the LB and Kubo formalisms.
In Figure~ \ref {fig:kubo} we present the these results for a system 
of
size $16\times
16$. The electron-hole
symmetry is also preserved in the Kubo formalism, so the spin-Hall conductance
vanishes at half filling, as in the LB formalism.

In Ref. \onlinecite{Sinitsyn},  Sinitsyn {\it et.} al.  and Shen use the Kubo formula  
to compute the spin-Hall conductance analytically, when both the Rashba and Dresselhaus couplings
are considered. As in our case, they find that the spin-Hall conductance vanishes
when  the Rashba and Dresselhaus couplings have the same strength. 
The predicted value of the SHC, however, is a constant  $\pm e/8\pi$, depending on 
the ration $V_R/V_D$. In contrast,
in our numerical approach, the SHC is no longer a universal constant, 
but rather a function of the 
Fermi energy and  of the Rashba/Dresselhaus coupling
strengths. 

\section{Conclusions}

In this work we have investigated the longitudinal and spin-Hall 
conductances of a two dimensional electronic system with Rashba and 
Dresselhaus spin-orbit coupling in the framework of a tight binding 
approximation.
For the main part of the work we have used Landauer-B\" uttiker 
formalism combined with Green's function approach to study the 
efect of spin-orbit coupling and disorder on $G_L$ and $G_{sH}$.
Our results for the spin Hall conductance, as function of Fermi energy and  
disorder strength, in the case when Dresselhaus coupling is neglected, agree 
with the results of  Ref. \onlinecite{Nikolic} which is a special case 
of the present model. 

We have also computed the Fermi energy dependence 
of longitudinal and spin-Hall conductances in the Kubo formalism. 
The good agreement found between the two sets of conductances computed 
in LB and Kubo formalisms  strengthens the
assumption that the spin-Hall effect is a bulk property of the
system. However, further studies are needed in order to clarify the role of terminals. 
For example, one can investigate the  scaling of spin-Hall conductance as function of
the system size, both in the Landauer-B\" uttiker and the Kubo
formalisms.

\begin{acknowledgments}
We gratefully acknowledge the financial support provided by the
Department of Energy, grant no. DE-FG02-01ER45897.
\end{acknowledgments}

\appendix
\section{Comparison with the Kubo formalism }

In this Appendix we present the derivation for the Kubo formula used
for computing the longitudinal and spin-Hall conductances.

\begin{figure}[h]
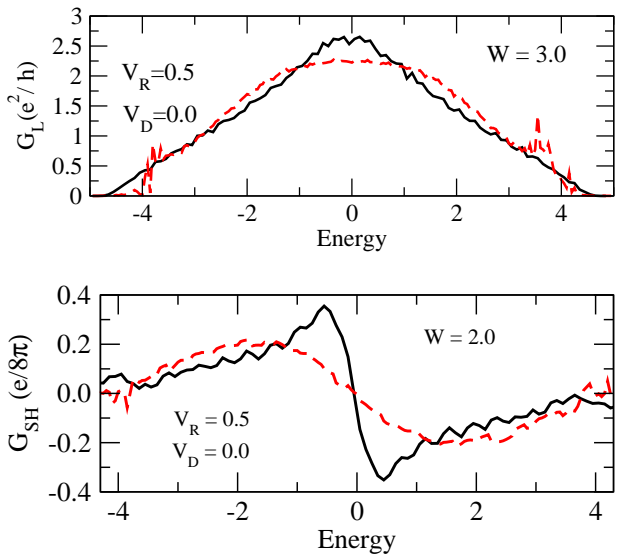

\begin{center}
\includegraphics[width=3.2in]{conductance_kubo.eps}
\end{center}
\begin{center}
\includegraphics[width=3.2in]{spin_hall_conductance_kubo.eps}
\end{center}
\caption{(Color online) Longitudinal (upper panel) and spin-Hall conductance (lower panel)
for a system size of  $16\times 16$. Solid lines represent results obtained using
Kubo formalism while the dashed lines are obtained using Landauer-B\" uttiker
method. Results are average over 500 samples. Energy is measured in units of $t$.}
\label{fig:kubo}
\end{figure}
In terms of single particle states, the longitudinal conductivity
can be computed from the general Kubo formula:
\begin{equation}
\sigma_{L}\left({\mathbf r}, {\mathbf r}' \right) = -i\,\hbar \sum_{n,n'}\frac{P_{n'} -P_{n}}{E_{n'}-E_{n}}
    \frac{\left< n' \left|{j_{x}}({\mathbf r})\right | n \right>
    \left< n \left|{v_{x}}({\mathbf r}')\right | n'\right>}
    {E_{n'}-E_{n}+i\, \hbar 0^{+}}.
\label{eq:longitudinal_conductance}
\end{equation}
Similarly, for the spin-Hall conductance we write:
\begin{equation}
\sigma_{sH}\left({\mathbf r}, {\mathbf r}' \right) =\sum_{n,n'}\frac{P_{n'} -P_{n}}{E_{n'}-E_{n}}
\frac{Im \left< n' \left| j_{x}^{z}({\mathbf r})\right | n \right> 
\left< n \left| v_y({\mathbf r}')\right | n'\right>}
{E_{n'}-E_{n}+i\, \hbar 0^{+}}. \label{eq:kubo_spin_hall_conductance}
\end{equation}
The single-particle states are constructed from the site orbitals as
$b_{n}^{\dagger}=\sum_{i, \alpha} \psi_n\left(i,\alpha\right)
c_{i\alpha}^{\dagger}$. Operator $b_{n}^{\dagger}$ stands for the
creation of a single particle state $\left | n\right >$ from the
one-electron wave functions, $\psi_n\left(i,\alpha\right)$. The wave
functions $\psi_n\left(i,\alpha\right)$ and the corresponding
eigen-energies $E_{n}$ can be easily obtained by solving the
eigen-value problem for the Hamiltonian
(\ref{eq:tight_binding_hamiltonian}).

The velocity operator is defined by the commutator: $i\, \hbar
{\mathbf v} = \left[ {\mathbf r}, H \right ]$, while the spin
current is given in terms of the anticommutator between the velocity
operator and Pauli matrix $\sigma_z$: $j_x^{z} =\hbar/4\left\{
\sigma_z, v_x \right\}$. A simple quantum mechanics calculation
gives for the current and for the spin-current operators
the following expressions:
\begin{equation}
\left < n\left | {\mathbf v}\right | n'\right >  = \frac{1}{i\, \hbar} \sum_{i,j,\alpha, \beta}
    \psi_{n}^{*}\left (i, \alpha \right )
    \left[ \left( {\mathbf r}_{i} -{\mathbf r}_{j}\right)  H_{ij}^{\alpha, \beta}    \right ]
    \psi_{n}\left (j, \beta \right )
\label{eq:current}
\end{equation}
\begin{equation}
\left < n\left | {\mathbf j}^{z}\right | n'\right >  = \frac{e}{ 4\, i} \sum_{i,j,\alpha, \beta}
    \psi_{n}^{*}\left (i, \alpha \right )
    \left[ \left( {\mathbf r}_{i} -{\mathbf r}_{j}\right)  \tilde{H}_{ij}^{\alpha, \beta}  \right ]
    \psi_{n}\left (j, \beta \right ).
\label{eq:spin_current}
\end{equation}
In Eq. (\ref{eq:spin_current}), $\tilde{H} = \left\{\sigma_z\otimes
{\mathbf 1}, H \right \}$.

At $T=0K$, when the Fermi function derivative is approximated by a
delta function, we write:
\begin{eqnarray}
\frac{P_{n'}-P_{n}}{E_{n'}-E_{n}}  &= &\int dE\,\frac{\partial f (E)}{\partial E}\, \delta(E_{n}-E) \nonumber\\
                                &= &-\delta(E_n - E_F).\label{eq:distribution}
\end{eqnarray}
Incorporating Eqs. (\ref{eq:current}), (\ref{eq:spin_current}),
(\ref{eq:distribution})
 in Eqs.~(\ref{eq:longitudinal_conductance})
and (\ref{eq:kubo_spin_hall_conductance}) a simple expression for
longitudinal and spin-Hall conductance in terms of single-particle
wave functions and eigen-energies is obtained. We note that only
terms at the Fermi levels give contributions to the longitudinal
conductance therefore we keep only the delta function part from
$(E_{n'}-E_{n}+i\,\hbar 0^{+})^{-1}$, in
Eq.~(\ref{eq:longitudinal_conductance}). In this respect,  the
longitudinal conductance is a sum of weighted delta functions which
have to be broadened into functions having a finite width (for
example a Lorentzian). When the spin-Hall conductance is computed,
the principal value of $(E_{n'}-E_{n}+i\,\hbar 0^{+})^{-1}$ is
needed in Eq.~(\ref{eq:kubo_spin_hall_conductance}).

\end{document}